\documentclass[dvips]{arxstspdf}
\usepackage{flushend}
\usepackage{stfloats}
\volume{24}
\issue{4}
\pubyear{2009}
\firstpage{489}
\lastpage{502}
\doi{10.1214/09-STS297}

\makeatletter
\DeclareMathAlphabet\mathcaligr{OMS}{cmsy}{m}{n}
\def\be{\begin{eqnarray}}
\def\ee{\end{eqnarray}}

\def\bse{\begin{eqnarray*}}
\def\ese{\end{eqnarray*}}

\def\log{\operatorname{log}}
\def\cov{\operatorname{cov}}

\def\trans{^\mathrm{T}}

\def\pr{\operatorname{pr}}
\def\wh{\widehat}
\def\n{N}
\def\N{\mathcaligr{N}}
\def\H{\mathcaligr{H}}
\def\di{^{\mathrm{di}}}
\newcommand{\eqref}[1]{(\ref{#1})}
\renewcommand{\citet}[1]{\citeauthor{#1} (\citeyear{#1})}
\renewcommand{\citep}[1]{(\citeauthor{#1}, \citeyear{#1})}
\makeatother

\begin{document}
\begin{frontmatter}

\title{Analysis of Case-Control Association Studies: SNPs, Imputation
and Haplotypes}
\runtitle{SNPs, Haplotypes and Imputation}

\begin{aug}
\author[a]{\fnms{Nilanjan} \snm{Chatterjee}\ead[label=e1]{chattern@mail.nih.gov}\corref{}},
\author[b]{\fnms{Yi-Hau} \snm{Chen}\ead[label=e2]{yhchen@stat.sinica.edu.tw}},
\author[c]{\fnms{Sheng} \snm{Luo}\ead[label=e3]{Sheng.T.Luo@uth.tmc.edu}} \and
\author[d]{\fnms{Raymond J.} \snm{Carroll}\ead[label=e4]{carroll@stat.tamu.edu}}
\runauthor{Chatterjee, Chen, Luo and Carroll}

\affiliation{National Cancer Institute, Academia Sinica, University of
Texas School of Public Health, Texas A\&M University}

\address[a]{Nilanjan Chatterjee is Chief and Senior Investigator, Division of Cancer
Epidemiology and Genetics, National Cancer Institute, NIH, DHHS,
Rockville Maryland 20852, USA \printead{e1}.}
\address[b]{Yi-Hau Chen is Research Fellow, Institute of Statistical
Science, Academia Sinica, Taipei 11529, Taiwan, ROC \printead{e2}.}
\address[c]{Sheng Luo is Assistant Professor, Division of Biostatistics,
University of Texas School of Public Health, Houston, Texas 77030, USA
\printead{e3}.}
\address[d]{Raymond J. Carroll is Distinguished Professor, Department of
Statistics, Texas A\&M University, College Station, Texas 77843-3143,
USA \printead{e4}.}

\end{aug}

%
\begin{abstract}
Although prospective logistic regression is the standard\break method of
analysis for case-control data, it has been recently noted that in
genetic epidemiologic studies one can use the ``retrospective''
likelihood to gain major power by incorporating various population
genetics model assumptions such as Hardy--Weinberg-Equilibrium (HWE),
gene--gene and gene--environment independence. In this article we
review these modern methods and \mbox{contrast} them with the more
classical approaches through two types of applications (i)
association tests for typed and untyped single nucleotide
polymorphisms (SNPs) and (ii) estimation of haplotype effects and
haplotype--environment interactions in the presence of
haplotype-phase ambiguity. We provide novel insights to existing
methods by construction of various score-tests and
pseudo-likelihoods. In addition, we describe a novel two-stage
method for analysis of untyped SNPs that can use any flexible
external algorithm for genotype imputation followed by a powerful
association test based on the retrospective likelihood. We
illustrate applications of the methods using simulated and real
data.
\end{abstract}
%
%

\end{frontmatter}
%
\section{Introduction} \label{sec:sec1}

Case-control study designs are now widely used to study the role
of genetic susceptibility in the etiology of rare complex
diseases. Typically, a case-control study involves recruiting all
or a large fraction of the diseased subjects (cases) that arise in
an underlying study base and then sampling a comparable number of
healthy subjects (controls), ideally from the exact same study
base, and possibly matched with the cases by some
socio-demographic characteristics such as race, age and gender.
Biological samples and questionnaire data collected on the sampled
subjects are then used to determine their genetic susceptibility,
such as SNP genotypes and history of some nongenetic
(environmental) exposures. For rare diseases such as cancers,
case-control studies are cost-efficient compared to a
cross-sectional or prospective cohort studies because they
dramatically reduce the number of nondiseased subjects to study.

In general, the standard method for analysis of case-control data
is the prospective logistic regression ignoring the retrospective
nature of the underlying design. The validity of this approach
relies on the classic results by \citeauthor{Cornfield1956} (\citeyear{Cornfield1956}) who showed
the equivalence of prospective- and retrospective-odds ratios. The
efficiency of the approach was established in two other classic
papers by \citeauthor{Andersen1970JRSSB} (\citeyear{Andersen1970JRSSB}) and
\citet{Prentice1979Biometrika}, who showed that the prospective
analysis of case-control data yields the proper maximum-likelihood
estimates of the odds ratio parameters of the logistic model under
a ``semiparametric'' setup that allows the distribution of the
underlying covariates to remain completely unrestricted. More
recently, it has been shown that even in the presence of missing
data and measurement error in covariates, the ``prospective''
treatment of case-control data can yield proper maximum-likelihood
estimates as long as the distribution of the underlying covariates
is allowed to remain unrestricted \citep{Roeder1996JASA}.

A special feature for studies in genetic epidemiology is that it
is often reasonable to assume certain models for the population
distribution of the genetic and environmental covariates of
interest. The Hardy--Weinberg-Equilibrium (HWE) law, for example,
which implies a simple relationship between \textit{allele} and \textit{genotype} frequencies at a given chromosomal locus, is a natural
model for a random mating, large, stable population in the absence
of new genetic mutations, inbreeding and selective survivorship
among genotypes (see \cite{Hartl2007}, Chapter 3). Genes which are
physically apart and hence are not expected to be in linkage
disequilibrium (LD) are also expected to be independently
distributed in a homogeneous population. It is often also natural
to assume a subject's genetic susceptibility, a factor which is
determined at birth, is independent of his/her subsequent
environmental exposures. A~pertinent question then is what is the
most appropriate method for analysis of case-control data in
genetic epidemiology where some natural model assumptions exist
for the distribution of genetic and environmental factors in the
underlying population.

We will assume data on some genetic ($G$) and environmental ($E$)
exposures are collected in a case-control study involving $N_0$
controls ($D=0$) and $N_1$ cases ($D=1$). If one ignores the
retrospective nature of the case-control design, one can conduct
inference based on the prospective-likelihood
\begin{equation}\label{eq:PL}
L^P=\prod_{i=1}^{N}\operatorname{pr}(D_i|G_i,E_i),
\end{equation}
where $N=N_1+N_0$. The fundamental likelihood for case-control data,
however, known as the ``retrospective'' likelihood, is given by
\begin{equation}\label{eq:RL}
L^R=\prod_{i=1}^{N}\operatorname{pr}(G_i,E_i|D_i).
\end{equation}

In the absence of any missing data, it is evident from the
classical theory that the prospective-likelihood (\ref{eq:PL})
provides a valid way of testing and estimation of the odds ratio
association parameters of the underlying logistic regression
model. In fact, the prospective-likelihood yields the same
maximum-likelihood estimates for the odds ratio association
parameters that could be obtained by maximization of the proper
retrospective likelihood (\ref{eq:RL}) while allowing
$\operatorname{pr}(G,E)$, the joint distribution of $G$ and $E$, to remain
completely non-parametric. Under constraints on $\operatorname{pr}(G,E)$,
however, the retrospective likelihood would not yield the same
maximum-likelihood estimator as that from the prospective
likelihood. More importantly, the retro\-spective-likelihood can
exploit various population genetics model assumptions such as
HWE, gene--gene and gene--environment independence to gain major
efficiency over the prospective-likelihood for inference on
various association and interaction parameters. At the same time,
if the underlying model assumptions are violated, then the use of
the retrospective likelihood can lead to serious bias for both
testing and estimation procedures. In the presence of missing
data, a further complication is that the use of the prospective
likelihood may not be even strictly valid in certain settings,
such as that described in Section \ref{sec:sec4} for estimation of haplotype
effects, where for the purpose of identifiability $L^P$ also
requires some modeling assumptions, thus destroying its
equivalence with $L^R$ that is known to hold under unspecified
covariate distribution. Thus, to date, a debate remains about the
most appropriate method of analysis of case-control studies in
genetic epidemiology.

In this article we will review some modern developments for
analysis of case-control studies in genetic epidemiology using the
prospective- and\break retrospective-likelihoods. We will describe the
methods primarily through two different types of applications: (a)
association testing for genotyped and imputed single nucleotide
polymorphisms (SNP) (Sections \ref{sec:sec2} and \ref{sec:sec3})
and~(b) estimation of haplotype effects and haplotype--environment
interactions in the presence of phase ambiguity (Section
\ref{sec:sec4}). In each section we aim to provide new intuitive
insights into the alternative methods by constructions of various
score tests (Sections \ref{sec:sec2} and~\ref{sec:sec3}) and
pseudo-likelihoods (Section \ref{sec:sec4}). As a byproduct, in
Section \ref{sec:sec3} we also propose a novel ``retrospective''
method for association testing for untyped SNPs which can easily
use any external algorithm for imputation of genotypes. In each
section we will use numerical examples to illustrate the bias and
efficiency trade-off \mbox{between} the alternative methods. We will
conclude the article with a discussion and recommendations for
practical data analysis.
\section{Association Analysis for Single Nucleotide Polymorphisms
(SNPs)}\label{sec:sec2}

\subsection{The Prospective Approach}
The genotype information for an individual SNP in a case-control
study can be represented by the $2\times3$ contingency table
defined by cross-tabulation of case-control and genotype status.
Let $D$ be the indicator of case ($D=1$) or control ($D=0$) status
and let $G$ be the number of minor alleles carried by an individual
($G=0,1,2$). Let $n_{dg}$ denote the number of subjects with
genotype $G=g$ and disease status $D=d$ observed in the case-control
sample. Suppose we are interested in testing the association of
the disease outcome with a SNP-genotype using a population logistic
regression model of the form
\begin{equation}\label{eq:logistic} \operatorname{pr}(D=1|G)=\frac{\exp\{
\alpha+\beta\trans m(G)\}}{1+\exp\{\alpha+\beta\trans
m(G)\}},
\end{equation}
where the function $m(\cdot)$ is chosen in a suitable way to
reflect an assumed mode of genetic effect. If, for example, $G$
denotes the count for the minor allele at a SNP locus, then one
can chose $m(G)=G$, $m(G)=I(G\geq1)$ or $m(G)=I(G=2)$ to model
the effect of the minor allele as additive (in the logistic
scale), dominant or recessive. One can also consider a saturated
model by allowing $m(G)$ to be a vector of two dummy variables
associated with heterozygous ($G=1$) and homozygous variant
($G=2$) genotypes and $\beta$ to be the corresponding
log-odds-ratios. The prospective analysis of case-control data
yields an asymptotically unbiased estimate for the genotype-odds-ratio
parameters~$\beta$, but not for the intercept parameter $\alpha$.

The score function for $\beta$ under the prospective-likelihood
(\ref{eq:PL}) can be written as
\[
U_{PL}=\sum_{i=1}^{N_1+N_0}m(G_i)\{D_i-\operatorname{pr}(D=1|G_i)\}.
\]
Under the null hypothesis, $\beta=0$, we can estimate
$p=\operatorname{pr}(D=1|G_i)$ as $\widehat p = N_1/(N_1+N_0)$ since under
that hypothesis, $G$ does not influence $D$. Then the score
function can be written as
\begin{eqnarray*}
U_{PL}^{0}&=&\sum_{i=1}^{N_1}m(G_i)-\frac{N_1}{N_1+N_0}\sum
_{i=1}^{N_1+N_0}m(G_i)\\
&=&\frac{N_1N_0}{N_1+N_0}\Biggl\{\frac{1}{N_1}\sum
_{i=1}^{N_1}m(G_i)-\frac{1}{N_0}\sum_{i=N_1+1}^{N_1+N_0}m(G_i)\Biggr\},
\end{eqnarray*}
which is proportional to the difference between the empirical
means of $m(G)$ in the cases ($D=1$) and in the controls ($D=0$).
We suppose without loss of generality that the indices for the
cases are $\{i=1,\ldots,N_1\}$ and those for the controls are
$\{i=N_1+1,\ldots,N_1+N_0\}$. If, for example, we assume $m(G)=G$,
that is, the additive effect, then $U_{PL}^{0}$ corresponds to the
numerator of the Cochran--Armitage trend test
(\citeauthor{VanBelle2004BiostatisticsBook}, \citeyear{VanBelle2004BiostatisticsBook}, Chapter 7) that is widely
used for single-SNP association testing. More generally, a
``prospective'' score-test can be constructed under any genetic
model based on $U_{PL}^{0}$ and its variance under the null
hypothesis of no association that be estimated by
\[
V^{0}_{PL}=\frac{N_1N_0}{N_1+N_0}V_{m(G)},
\]
where $V_{m(G)}$ is the pooled-sample variance of $m(G_i)$.

\subsection{Retrospective Approach}
The retrospective likelihood, $L^{R}$, for the genotype data of a
single-SNP can be written as the product of two sets of
multinomial probabilities:
\[
L^{R}=L_1\times
L_0=\prod_{g=0}^2p_{1g}^{n_{1g}} \times
\prod_{g=0}^2p_{0g}^{n_{0g}},
\]
where
$p_{dg}=\operatorname{pr}(G=g|D=d)$, $d=0$ and $1$, denotes the population
genotype frequencies for the controls and the cases, respectively.
Given the genotype probabilities for the controls, we can
characterize the genotype probabilities for the cases according to
the formula \citep{Satten1993JASA}
\begin{equation}\label{eq:gp1}
p_{1g}=\frac{\psi_{g}(\beta)p_{0g}}{\sum_{g=0}^2\psi_{g}(\beta)p_{0g}},
\end{equation}
where $\psi_{g}(\beta)$ denotes the odds ratio associated with the
genotype $G=g$ as specified by the logistic model
(\ref{eq:logistic}). Thus, the retrospective likelihood can be
parameterized in terms of the genotype probabilities of the
controls and the disease-odds-ratio parameters $\beta$. The
maximization of the retrospective likelihood $L^R$, without
imposing any further constraints on the genotype probabilities for
the controls, will lead to the same estimator for $\beta$ that
would be obtained by maximization of $L^P$
\citep{Prentice1979Biometrika}. In fact, it can be shown that the
retrospective- and prospective-profile likelihoods of $\beta$
become identical after maximization of the corresponding
likelihoods with respect to the associated nuisance parameters
\citep{Roeder1996JASA}. Thus, the associated tests, including
score-, Wald- and likelihood-ratio tests, are identical under the
retrospective and prospective likelihoods.

Now suppose we are willing to assume that HWE holds in the underlying
population and that the disease is rare so that HWE also holds
approximately in the control population. In the retrospective
likelihood $L^R$, we can write the genotype probabilities for the
controls as a function of the frequency, $f$, of the minor allele
as
\begin{eqnarray*}
p_{00}(f)&=&(1-f)^2,\qquad p_{01}(f)=2f(1-f),\\
p_{02}&=&f^2.
\end{eqnarray*}
It is easy to show that the score function for $\beta$ associated
with the retrospective likelihood can be written as
\[
U_{RL}=\sum_{i=1}^{N_1}[m(G_i)-E_{\mathrm{HWE},f}\{m(G)|D=1\}
],
\]
which under the null hypothesis of no association reduces to
\begin{equation}\label{eq:RScore}
U_{RL}^{0}=\sum_{i=1}^{N_1}[m(G_i)-E_{\mathrm{HWE},f}\{m(G)\}
].
\end{equation}
Moreover, under the null hypothesis, the allele frequency $f$ can
be substituted for by its maximum-likelihood estimate
\begin{equation}
\widehat{f} =\frac{n_{+1}+2n_{+2}}{2N}, \label{eq:qrjc01}
\end{equation}
where $n_{+g}$
denotes the frequency for genotype $G=g$ in the pooled sample of
cases and controls. Thus, $U_{RL}^{0}$ corresponds to the
difference between the empirical mean of the function $m(G)$ in
cases and its expected value under HWE and the null hypothesis of
no association. In contrast, note that $U_{PL}^{0}$ corresponds
to the difference between the empirical mean of the function
$m(G)$ in cases and the empirical mean for the same function in
the controls. If the expectation in the retrospective score
function (\ref{eq:RScore}) is estimated empirically without
assuming HWE, then, as expected, it can be easily shown that the
retrospective and prospective scores are the same. If, however, we
assume HWE to evaluate the retrospective score function, then it
would have smaller variance than that for the prospective score.
In particular, this can be seen from the estimate of the variance
estimate $U_{RL}^{0}$ given by
\[
V^{0}_{RL}=N_1\biggl\{V_{m(G)}-\frac{N_1}{2N}\hat f(1-\hat f)C(\hat
f)C(\hat f)\trans\biggr\},
\]
where
\begin{eqnarray*}
C(f)&=&\cov_{\mathrm{HWE},f}\biggl\{m(G),
\frac{G-2f}{f(1-f)}\biggr\}\\
&=&\sum_g m(g)\frac{g-2f}{f(1-f)}p_{0g}(f).
\end{eqnarray*}
By the Cauchy--Schwarz inequality, $V^{0}_{RL}\ge V^{0}_{PL}$
asymptotically, implying that the retrospective
score test is asymptotically more powerful than its prospective
counterpart when the assumption of HWE is valid.

\citeauthor{Chen2007HumHered} (\citeyear{Chen2007HumHered}) compared the performance of 2 d.f.
Wald-tests of association based on the retrospective and
prospective likelihoods. They observed major gains in power for
the test based on the retrospective-likelihood for the detection
of nonmultiplicative effects, for example, recessive effects. Notice
that if we assume an additive model, that is, $m(G)=G$, then the
prospective and retrospective score functions $U_{RL}^{0}$ and
$U_{PL}^{0}$ become identical because in this case
$E_{\mathrm{HWE},\widehat f}\{m(G)\}=2\widehat f=\sum_{i=1}^{N}G_i/N$. The
larger the departure of the effect of a SNP from the additive
form, the greater the gain in efficiency for the retrospective
method. Application of retrospective methods for association
testing, however, requires caution \mbox{because} of their sensitivity to
the underlying model assumption. In particular, it can be seen
from the formula of $U_{RL}^{0}$ that the unbiasedness of that
score \mbox{function} crucially depends on the assumption of HWE being
correct for the underlying population. \citeauthor{Satten2004GenetEpi}~(\citeyear{Satten2004GenetEpi})
and \citeauthor{Chen2007HumHered} (\citeyear{Chen2007HumHered}) have noted that even modest violation
of HWE can cause serious inflation in Type-I error in association
tests based on the retrospective likelihood.

\subsection{Empirical-Bayes Methods}\label{subsec:2.3}
\citeauthor{Luo2009GeneticEpi} (\citeyear{Luo2009GeneticEpi}) considered an empirical-Bayes type
shrinkage estimation approach to develop a 2 d.f. single-SNP
association test that can gain power by exploiting the model
assumptions of HWE for the underlying population and yet is
resistant to bias when the model assumptions are violated. The
method involves estimation of genotype-specific disease odds ratio
parameters by data-adaptive ``shrinkage'' of a ``prospective''
model-free estimator that does not require the HWE assumption
toward a ``retrospective'' model-based estimator that directly
exploits the HWE constraints. The amount of\break ``shrinkage'' is
sample-size and data-adaptive, so that in large samples the method
has no bias whether the assumption of HWE holds or not and yet the
method can gain efficiency by shrinking the analysis toward HWE,
but only to the extent that the data validate the assumptions. In
what follows we provide some insight into the empirical-Bayes
method through the construction of a score-test. For numerical
illustration, however, we will focus on the Wald test as
originally developed in \citeauthor{Luo2009GeneticEpi} (\citeyear{Luo2009GeneticEpi}).

Let $\overline{m}(G)=(N_1+N_0)^{-1}\sum_{i=1}^{N_1+N_0}m(G_i)$\break
and $s^2_{\overline{m}(G)}=(N_1+N_0)^{-1}\sum_{i=1}^{N_0+N_1}\{
m(G_i)-\overline{m}(G)\}^2$
denote the sample mean and variance for the function $m(G)$, respectively.
Further, let $\widehat\tau=\overline{m}(G)-E_{\widehat f,\mathrm{HWE}}m(G)$ denote
the difference between the empirical and expected means of $m(G)$
when the latter quantity is computed assuming HWE and under the
estimate of allele frequency $\widehat f$ given in (\ref{eq:qrjc01}). %
Intuitively, $\widehat\tau$ can be viewed as an estimate of the
bias in estimation of the population mean of $m(G)$ under the
assumption of HWE. An empirical-Bayes type score function can be
now defined as
\begin{equation}\label{eq:EBScore}
U_{\mathit{EB}}^{0}=\sum_{i=1}^{N_1}[m(G_i)-E_{\mathit{EB}}\{m(G)\}],
\end{equation}
where $E_{\mathit{EB}}\{m(G)\}$ is the empirical-Bayes estimate
for the mean of the function $m(G)$ under $H_0$, given by
\begin{eqnarray*}
E_{\mathit{EB}}\{m(G)\}&=&\frac{s^2_{\overline{m}(G)}/
N}{s^2_{\overline{m}(G)}/N+\widehat\tau^2}
E_{\mathrm{HWE},\widehat f}\{m(G)\}\\
&&{}+\frac{\widehat\tau
^2}{s^2_{\overline{m}(G)}/N+\widehat\tau^2}\overline{m}(G).
\end{eqnarray*}
Thus, $E_{\mathit{EB}}\{m(G)\}$ corresponds to a weighted
average of the empirical mean of $m(G)$ and its expected mean
under HWE, with the weights defined by an estimate of the bias for
the estimate of the population mean of $m(G)$ under HWE and an
estimate of the variance of the empirical mean of $m(G)$. As
$\widehat\tau^2$ decreases, that is, the evidence of bias due to the
violation of HWE becomes smaller, $E_{\mathit{EB}}\{m(G)\}$
gives more weight to the more precise HWE-based estimator of the
population mean of $m(G)$. Conversely, as
$s^2_{\overline{m}(G)}/N$ decreases, that is, the sample mean of $m(G)$
becomes more precise, then $E_{\mathit{EB}}\{m(G)\}$ puts more
weight to the robust model-free estimator $\overline{m}(G)$. The
original perspective for constructing such\break weighted combinations
of model-based and model free estimators from an empirical-Bayes
point of view can be found in \citeauthor{Mukherjee2008Biometrics} (\citeyear{Mukherjee2008Biometrics}).
Simple methods for variance estimation for such estimators have
been also described in that article.
\subsection{The Cancer Genetics Markers of Susceptibility (CGEMS)
Study}\label{sec:CGEMS}

We illustrate the performance of alternative 2 d.f. single SNP
association tests using data from the Cancer Genetics Markers of
Susceptibility (CGEMS) study (\citeauthor{Yeager2007NatGenet}, \citeyear{Yeager2007NatGenet};
\citeauthor{Hunter2007NatGenet}, \citeyear{Hunter2007NatGenet};\break
\citeauthor{Thomas2008NatGenet}, \citeyear{Thomas2008NatGenet}), an NCI enterprize
initiative to conduct multistage whole-genome association studies
to identify susceptibility genes giving rise to increased risks of
prostate and breast cancers. In this article we will focus on data
from the initial scan for the prostate cancer study, involving
genotype data on about $550\mbox{,}000$ SNPs from $1172$ cases and $1157$
controls. The details of the CGEMS study design and the results from
the initial scan and subsequent replication studies can be found at
the web site \url{https://caintegrator.nci.nih.gov/cgems/}.

\begin{table}
\caption{The empirical proportions of significant SNPs detected by
different methods at different nominal significance levels in the
CGEMS prostate cancer study}\label{Tab:observedPval}
\begin{tabular*}{\columnwidth}{@{\extracolsep{\fill}}lccc@{}}
\hline
$\bolds{\alpha}$ & \textbf{Prospective} & \textbf{Retrospective} & \textbf{Empirical-Bayes} \\
\hline
5e--2 & 5.01e--2 & 5.66e--2 & 4.49e--2 \\
1e--2 & 0.98e--2 & 1.43e--2 & 0.87e--2 \\
1e--3 & 1.05e--3 & 3.85e--3 & 1.00e--3 \\
1e--4 & 1.27e--4 & 2.24e--3 & 1.31e--4 \\
1e--5 & 2.67e--5 & 1.76e--3 & 3.34e--5 \\
1e--6 & 2.22e--6 & 1.47e--3 & 4.45e--6 \\
\hline
\end{tabular*}
\end{table}

We consider $449\mbox{,}698$ SNPs from 22 nonsex chromosomes with minor
allele frequencies larger than $0.05$.
Table \ref{Tab:observedPval} displays the empirical proportions of
the number of SNPs that are found to be significant at different
nominal significance levels using 2 d.f. tests based on three
different methods: (a) prospective, (b) retrospective and (c)
empirical-Bayes [see \citeauthor{Luo2009GeneticEpi} (\citeyear{Luo2009GeneticEpi}) for more details].
For a well-designed study and a robust analytic method, the
empirical proportions are expected to be fairly close to the
nominal significant levels, given that the vast majority of the SNPs
are likely to be not associated with the disease. In
Table \ref{Tab:observedPval}, we observe that the empirical
proportions of significant SNPs found by the prospective method
closely follows the nominal significance levels. In contrast, the
corresponding proportions for the retrospective test deviate
severely from the nominal values in the range of $\alpha\leq
10^{-3}$, indicating significantly inflated type-I error due to
the violation of HWE for many SNPs. The last column of
Table \ref{Tab:observedPval} shows that the empirical-Bayes
procedure essentially corrects for all the bias of the
retrospective method due to the violation of the HWE assumption.

Next, we conducted a simulation study to investigate the
performance of various tests in ranking a true susceptibility
locus in a genome-wide association study (GWAS) that include
hundreds of thousands of ``null'' SNPs. To generate realistic
linkage disequilibrium patterns, we simulated GWAS data mimicking
the CGEMS study itself. Given minor allele frequency among
controls and the disease-genotype odds ratio parameters for a
chosen susceptibility locus, we simulate genotype data at that
locus for the cases and controls separately from the corresponding
multinomial distributions. Given the genotype data at the
susceptibility locus for a case or a control, we simulate genotype
data for the remainder of the SNPs by assigning the whole genotype
profile for a randomly selected subject from the controls of the
CGEMS study who have the same genotype data at the given
susceptibility locus as the sampled subject in our simulation
study. This algorithm, as originally described by
\citeauthor{Yu2009GeneticEpi} (\citeyear{Yu2009GeneticEpi}), assumes that given the genotypes for the
susceptibility locus, the risk of the disease is independent of
all the remaining SNPs. We simulated $50$ data sets with
approximately $550$ cases and $550$ controls. For each data set
we tested for association for each of the approximately 450,000
SNPs using the prospective, retrospective and empirical-Bayes
methods. The rank of the disease-associated SNP is obtained by
sorting all the $p$-values in ascending order.

\begin{table}[b]
\caption{Simulated median ranks of a true susceptibility SNP with
a recessive effect and log-odds-ratio value of $\log(3)$ for
alternative tests. The results are based on $50$ simulated
datasets, each of which has approximately $550$ cases and $550$
controls and 450,000 SNPs. MAF: minor allele frequency}\label{Tab:rank_permute}
\begin{tabular*}{\columnwidth}{@{\extracolsep{\fill}}lccc@{}}
\hline
\textbf{MAF} & \textbf{Prospective} & \textbf{Retrospective} & \textbf{Empirical-Bayes} \\
\hline
0.1 & 112163 & 8117 & 44319 \\
0.2 & \phantom{00}1888 & \phantom{0}203 & \phantom{000}52 \\
0.3 & \phantom{000}656 & \phantom{0}210 & \phantom{000}27 \\
0.4 & \phantom{0000}15 & \phantom{00}82 & \phantom{0000}2\\
\hline
\end{tabular*}
\end{table}

Table \ref{Tab:rank_permute} displays the median ranks obtained by
three methods for a true disease-associated SNP that has a
recessive effect with a log-odds-ratio of $\beta=\log(3)$. As
expected, the ranks of all tests decrease as the minor allele
frequency increases. Comparing the ranks of different tests at a
specific minor allele frequency, we can see that the standard
prospective method generally has the lowest power in the sense
that it assigns much higher rank to the susceptibility SNP than
the two other tests. When minor allele frequency is $0.1$, we
observe that the pure retrospective method performs the best in
the sense that it assigns the lowest rank to the susceptibility
SNPs among all the methods. In contrast, when minor allele
frequency is greater than or equal to $0.2$, we observe that the
empirical-Bayes procedure assigns considerable lower rank to the
susceptibility SNP than the pure retrospective method.
Intuitively, the results can be explained from the fact that the
retrospective method yields low $p$-values for many null SNPs due to
the violation of the HWE assumption (see Table \ref{Tab:observedPval}) and thus
dilutes the rank of the real susceptibility SNP.
\section{Association Analysis for Imputed SNPs}\label{sec:sec3}

The forms of the prospective- and retrospective-scores suggest how
they can be modified easily for SNPs that may not have been
directly genotyped, but can be ``imputed'' conditional on
neighboring SNPs and estimates of linkage disequilibrium from
HapMap or other similar databases. Let $\N(G)$ denote the
neighboring genotype information for an untyped SNP-locus with
unobserved genotype $G$. The prospective score for such an untyped
SNP can be defined by taking the conditional expectation of the
``complete data'' score function $U_{PL}^{0}$ given the observed
data, that is, the neighboring genotype information. More formally, the
prospective score for an untyped SNP can be written as
\begin{eqnarray}
U_{PL}^{0u}&=&\frac{N_1N_0}{N_1+N_0}\Biggl[\frac{1}{N_1}\sum
_{i=1}^{N_1}E\{m(G)|\N(G_i)\}
\nonumber
\\[-8pt]
\\[-8pt]
\nonumber
&&\qquad{}\hspace*{22pt}-\frac{1}{N_0}\sum
_{i=1}^{N_0}E\{m(G)|\N(G_i)\}\Biggr],
\end{eqnarray}
where the conditional expectations are taken with respect to a
suitable imputation model such as those described by
\citeauthor{Nicolae2006GeneticEpi} (\citeyear{Nicolae2006GeneticEpi}),
\citeauthor{Marchini2007NatureGenetics} (\citeyear{Marchini2007NatureGenetics})
and others. The retrospective score for an untyped SNP can be
similarly defined by the conditional expectation of the ``complete
data'' retrospective score function $U_{RL}^{0}$ given the
observed data $\N(G)$ in the form
\begin{eqnarray}
U_{RL}^{0u}&=&\sum_{i=1}^{N_1}[E\{m(G)|\N(G_i)\}
\nonumber
\\[-8pt]
\\[-8pt]
\nonumber
&&\quad{}\hspace*{3pt}-E_{\mathrm{HWE},f}\{m(G)\}].
\end{eqnarray}
Notice that in the retrospective score function, the contribution of
the term $E_{\mathrm{HWE},f}\{m(G)\}$ is a constant term given the
allele frequency $f$. The estimation of the allele frequency $f$ for an
untyped SNP, however, requires imputation. In particular, under the
``complete data'' model we can write the estimate of the allele
frequency under the null hypothesis of no association as
\[
\widehat f =\frac{\sum_{i=1}^{N_0+N_1}\{I(G_i=1)+2I(G_i=2)\}
}{2N}.
\]
Thus, given an imputation model, we can estimate the allele
frequency $f$ as
\begin{eqnarray}\label{eq:f}
\hspace*{14pt}\widehat f^{u}&=&\Biggl(\sum_{i=1}^{N_0+N_1}\operatorname{pr}\{G=1|\N(G_i)\}
\nonumber
\\[-8pt]
\\[-8pt]
\nonumber
&&{}\hspace*{30pt}+2\mbox {pr}\{G=2|\N(G_i)\}\Biggr)\Big/(2N).
\end{eqnarray}

We further need the variances for $U_{PL}^{0u}$ and $U_{RL}^{0u}$
under the null hypothesis to obtain the corresponding score tests.
The variance of $U_{PL}^{0u}$ can be estimated as
\[
V^{0u}_{PL}=\frac{N_1N_0}{N_1+N_0}V_{E\{m(G)|\N(G)\}},
\]
where $V_{E\{m(G)|\N(G)\}}$ is the pooled-sample variance of
$E\{m(G)|\N(G_i)\}$. The
prospective-score test is based on the test statistic given by
\[
(U_{PL}^{0u})\trans\{V^{0u}_{PL}\}^{-}U_{PL}^{0u},
\]
where the superscripts T and---denote transpose and generalized
inverse, respectively. Asymptotically, this statistic follows a
chi-squared distribution under the null hypothesis of $\beta=0$,
with the degrees of freedom given by the dimension of $m(G)$. The
variance of the retrospective score $U_{RL}^{0u}$, after adjusting
for the estimation of the allele frequency $f$ by $\widehat f$
given by (\ref{eq:f}), can be estimated by
\begin{eqnarray*}
V^{0u}_{RL}&=&{N_1}\biggl[V_{E\{m(G)|\N(G)\}}\\
&&{}\hspace*{17pt}+\frac{N_1}{2N}\biggl\{\frac
{V_{E\{G|\N(G)\}}}{2}C(\hat
f)C(\hat f)\trans\\
&&{}\hspace*{52pt}-QC(\hat f)\trans-C(\hat f)Q\trans\biggr\}\biggr],
\end{eqnarray*}
where
$Q$ is the pooled-sample covariance between $E\{m(G)|N(G_i)\}$ and
$E\{G|N(G_i)\}$.
The variance of $U_{RL}^{0u}$ can also be alternatively estimated by
the robust sandwich-type estimate given as
\[
V^{0u}_{PL}=\sum_{i=1}^{N_1+N_0}\tilde U_{RL,i}^{0u}(\tilde
U_{RL,i}^{0u})\trans,
\]
where the efficient score
\begin{eqnarray*}
\tilde U_{RL,i}^{0u}&=&D_i[E\{m(G)|\N(G_i)\}-E_{\mathrm{HWE},\widehat
f}\{m(G)\}]\\
&&{}-\frac{N_1}{2N}C(\widehat f)
[E\{G|\N(G_i)\}-2\widehat f].
\end{eqnarray*}
The retrospective-score test is then based on the test statistic
given by
\[
(U_{RL}^{0u})\trans\{V^{0u}_{RL}\}^{-}U_{RL}^{0u},
\]
which again follows a chi-squared distribution asymptotically
under the null hypothesis, with the degrees of freedom given by
the dimension of $m(G)$. In both the prospective- and
retrospective-score tests given above, we obtain the conditional
probability\break $\Pr\{G|\N(G_i)\}$ directly from some external
reference database, for example, HapMap, a strategy similar to the
proposal of \citeauthor{Nicolae2006GeneticEpi} (\citeyear{Nicolae2006GeneticEpi}).


\def\di{^{\mathrm{di}}}

We now demonstrate the potential power advantages that might be
achieved by
imputing the untyped SNP, using numerical studies following two
scenarios as in Tables 1 and 2 of \citeauthor{Nicolae2006GeneticEpi} (\citeyear{Nicolae2006GeneticEpi}). In
Scenario 1 the untyped SNP can be perfectly predicted by the
genotypes of the typed SNPs, namely, the $R_s^2=1$ (see Stram et al.,
2004, for a definition), while in Scenario 2 the untyped SNP is
moderately predicted by the genotypes of the typed SNPs with
$R_s^2=0.39$. The SNP profiles together with the haplotype
frequencies estimated from HapMap CEU samples in the two scenarios
are summarized in Tables \ref{Tab:scenario 1} and
\ref{Tab:scenario 2}. Also listed in Tables \ref{Tab:scenario 1} and
\ref{Tab:scenario 2} are the haplotype frequencies we actually used
to simulate the SNP data for the case-control sample, which
moderately deviate from those seen in the HapMap CEU sample to
reflect the potential discrepancy between the HapMap and study
samples. The haplotype pair for each person is generated according
to HWE.

\begin{table}[b]
\tabcolsep=0pt
\caption{The SNP profiles and haplotype frequencies for the region
considered in Scenario 1, where the untyped SNP can be perfectly
predicted by genotyped SNPs $A_1,\ldots, A_4$ ($R^2_s=1$). Also
listed are the haplotype frequencies estimated from the CEU sample
in HapMap. Part of the data are from Table 1 of
\protect\citeauthor{Nicolae2006GeneticEpi} (\protect\citeyear{Nicolae2006GeneticEpi})}\label{Tab:scenario 1}
\begin{tabular*}{\columnwidth}{@{\extracolsep{\fill}}lcc@{}}
\hline
\multicolumn{1}{@{}c}{\textbf{Haplotype of SNPs}} & &\multicolumn{1}{c@{}}{\textbf{Frequency}} \\
$\bolds{A_1\mbox{--}T\mbox{--}A_2\mbox{--}A_3\mbox{--}A_4}$ & \multicolumn{1}{c}{\textbf{Frequency}} &
\multicolumn{1}{c@{}}{\textbf{from HapMap}}
\\
\hline
1--0--0--0--0 & 0.158 & 0.058\\
0--1--0--1--0 & 0.400 & 0.300\\
1--1--0--1--0 & 0.050 & 0.050\\
1--1--1--0--1 & 0.358 & 0.558\\
0--1--1--0--1 & 0.022 & 0.017\\
1--1--0--0--1 & 0.012 & 0.017\\
\hline
\end{tabular*}
\end{table}
\begin{table}
\caption{The SNP profiles and haplotype frequencies for the region
considered in Scenario 2, where the untyped SNP is moderately
predicted by genotyped SNPs $A_1,\ldots, A_3$ ($R^2_s=0.39$). Also
listed are the haplotype frequencies estimated from the CEU sample
in HapMap. Part of the data are from Table 2 of
\protect\citeauthor{Nicolae2006GeneticEpi} (\protect\citeyear{Nicolae2006GeneticEpi})}\label{Tab:scenario 2}
\begin{tabular*}{\columnwidth}{@{\extracolsep{\fill}}lcc@{}}
\hline
\multicolumn{1}{@{}c}{\textbf{Haplotype of SNPs}} & & \multicolumn{1}{c}{\textbf{Frequency }}\\
$\bolds{A_1\mbox{--}T\mbox{--}A_2\mbox{--}A_3}$ & \multicolumn{1}{c@{}}{\textbf{Frequency}} &
\multicolumn{1}{c@{}}{\textbf{from HapMap}}
\\
\hline
0--0--0--0 & 0.088 & 0.058\\
0--0--1--1 & 0.027 & 0.017\\
0--1--0--0 & 0.302 & 0.342\\
0--1--1--0 & 0.008 & 0.008\\
1--0--1--0 & 0.242 & 0.142\\
1--0--1--1 & 0.333 & 0.433\\
\hline
\end{tabular*}
\end{table}

We simulated the case-control status by the logistic regression model
(\ref{eq:logistic}), where the genetic determinant $G$ is given by
the minor allele count of the untyped SNP, and the function
$m(\cdot)$ is given by the recessive, dominant or additive genetic
mode. The intercept $\alpha=-3.0$, which yields an overall disease
rate around 5\%. Each analysis is based on a case-control sample
with 1000 cases and 1000 controls. The simulation results are based
on 1000 (3000) repetitions for evaluation of test power (size). All
the tests are performed at a significance level of 0.01. The
score tests are performed using the correct genetic model, and the
retrospective-score tests are based on the robust sandwich-type
variance estimates; results based on model-based variance estimates
are quite similar and are omitted. When performing the prospective-
and retrospective-score tests with imputed genotypes for the untyped
SNP, we use the haplotype frequency estimates from\break HapMap to obtain
the conditional probabilities\break $\Pr\{G|\N(G_i)\}$, even though the
case-control sample is actually from a population with moderately
different haplotype frequencies. To see the degree of recovery of
missing information achieved by imputation, we also perform the
prospective- and\break retrospective-score tests based on the true
genotypes at the untyped SNP. In addition, we perform the
multimarker Hotelling's $T^2$ test based on genotypes at typed SNPs
(\citeauthor{Xiong2002AJHG}, \citeyear{Xiong2002AJHG};
\citeauthor{Chapman2003HumanHeredity}, \citeyear{Chapman2003HumanHeredity}), which is
equivalent to the prospective-score test derived from the logistic
regression model (\ref{eq:logistic}) with the covariates $m(G)$
given as the vector of genotypes for all the typed SNPs.

Results for this simulation study are presented in Tables
\ref{Tab:scenario 1 result} (Scenario 1) and \ref{Tab:scenario 2
result} (Scenario 2). It is seen that the score tests with imputed
genotypes have size matching reasonably well with the nominal value
of 1\%, even though the imputation is based on haplotype frequencies
that are obtained from the HapMap data and are different from the
true frequencies. From the results regarding power, we see that
imputing the untyped SNP in either the prospective- or the
retrospective-score test can achieve substantial power gains as
compared with the Hotelling's $T^2$ test based only on genotyped
SNPs. The relative power improvement gained by imputation can still
be quite remarkable even when the accuracy for predicting the
untyped SNP using the genotyped SNPs is only of a moderate level
(Scenario 2, where $R^2_s=0.39$). On the other hand, the prediction
accuracy does affect the degree of recovery of the missing
information that may be achieved by imputation: in Scenario 1, with
perfect prediction of the untyped SNP, the tests using imputed
genotypes do attain the full power we would obtain if the tests
were based on the true genotype of the untyped SNP. In Scenario 2,
with moderate prediction of the untyped SNP, imputation of the
untyped SNP can recover partial but not full power. It is worth
remembering that, with exact data, the retrospective-score test is
usually more powerful than the prospective-score under the dominant
or recessive model, and the two tests are essentially equivalent
under the additive model. Here we observe the same phenomena when
the prospective- and retrospective-score tests are based on imputed
genotypes.

\begin{table*}
\caption{Size/Power (\%) of the prospective- and retrospective-score
tests (significance $\textit{level}=0.01$) based on the imputed and true (in
parenthesis) genotypes at the untyped causal SNP, using SNP data
generated according to Table \protect\ref{Tab:scenario 1} (perfect
prediction). Also shown are results for the Hotelling's $T^2$ test
based only on genotypes at the typed SNPs. Results for power (size)
are based on 1000 (3000) simulated data sets}\label{Tab:scenario 1 result}
\begin{tabular*}{300pt}{@{\extracolsep{\fill}}lccc@{}}
\hline
& \multicolumn{1}{c}{\textbf{Prospective score}} & \multicolumn
{1}{c}{\textbf{Retrospective score}} & \multicolumn{1}{c@{}}{\textbf{Hotelling's} $\bolds{T^2}$} \\
$\bolds{\beta}$ & \multicolumn{1}{c}{\textbf{imputed (true)}} &
\multicolumn{1}{c}{\textbf{imputed (true)}} & \\
\hline
\multicolumn{1}{@{}l}{Recessive model} &&&\\
0 & 1.1 (1.1) & 1.1 (1.1) &\phantom{0}0.9\\
0.5 &26.1 (26.1) & 33.7 (33.7) & \phantom{0}3.6\\
0.6 & 40.1 (40.1) & 55.3 (55.3) & \phantom{0}5.6\\[5pt]
\multicolumn{1}{@{}l}{Dominant model}& & &\\
0 & 1.0 (1.3) & 1.0 (1.3) & \phantom{0}0.9 \\
0.3 &68.6 (68.6) & 72.9 (72.9) & 39.0\\
0.4 & 96.0 (96.0) & 96.7 (96.7) & 79.3\\[5pt]
\multicolumn{1}{@{}l}{Additive model}& & &\\
0 & 1.2 (1.2) & 1.2 (1.2) & \phantom{0}0.9\\
0.2 & 43.0 (43.0) & 43.0 (43.0) & 24.2\\
0.3 & 86.4 (86.4) & 86.4 (86.4) & 65.5\\
\hline
\end{tabular*}
\end{table*}

\begin{table*}[b]
\caption{Size/Power (\%) of the prospective- and retrospective-score
tests (significance $\textit{level}=0.01$) based on the imputed and true (in
parenthesis) genotypes at the untyped causal SNP, using SNP data
generated according to Table \protect\ref{Tab:scenario 2} (moderate
prediction). Also shown are results for the Hotelling's $T^2$ test
based only on genotypes at the typed SNPs. Results for power (size)
are based on 1000 (3000) simulated data sets}\label{Tab:scenario 2 result}
\begin{tabular*}{300pt}{@{\extracolsep{\fill}}lccc@{}}
\hline
& \multicolumn{1}{c}{\textbf{Prospective score}} & \multicolumn
{1}{c}{\textbf{Retrospective score}} & \multicolumn{1}{c@{}}{\textbf{Hotelling's} $\bolds{T^2}$} \\
$\bolds{\beta}$ & \multicolumn{1}{c}{\textbf{imputed (true)}} &
\multicolumn{1}{c}{\textbf{imputed (true)}} & \\
\hline
\multicolumn{1}{@{}l}{Recessive model} &&&\\
0 & 1.4 (1.2) & 1.2 (1.2) & \phantom{0}1.1\\
0.5 &42.6 (92.2) & 47.0 (97.6) & 17.6\\
0.6 & 59.4 (99.1) & 66.4 (99.9) & 24.9\\[5pt]
\multicolumn{1}{@{}l}{Dominant model}& & &\\
0 & 0.8 (1.1) & 0.9 (1.0) & \phantom{0}1.1\\
0.4 &48.5 (95.6) & 54.3 (98.2) & 23.8\\
0.5 & 71.6 (99.6) & 77.2 (100)\phantom{.} & 41.5\\[5pt]
\multicolumn{1}{@{}l}{Additive model}& & &\\
0 & 1.0 (1.3) & 1.0 (1.3) & \phantom{0}1.1\\
0.3 & 60.2 (97.6) & 60.1 (97.6) & 40.6\\
0.4 & 92.5 (99.9) & 92.4 (99.9) & 77.4\\
\hline
\end{tabular*}
\end{table*}

As we noted earlier, when exact genotype data are available, the
retrospective-score test is more sensitive to violation of the HWE
assumption than the prospective-score test; that is, the former is
usually biased while the latter still remains unbiased when HWE
does not hold. To assess the robustness properties for the
prospective- and retrospective-score tests with imputed genotype
data, we performed a further simulation study where the SNP
haplotypes are still given as in Tables \ref{Tab:scenario 1} and
\ref{Tab:scenario 2}, but the haplotype pair $H\di=(h_a,h_b)$ for
each person is given by the model with
$\Pr\{H\di=(h_a,h_b)\}=(1-\zeta)\theta_{a}\theta_{b}$ for $h_a\ne
h_b$ and $\Pr\{H\di=(h_a,h_b)\}=\zeta
\theta_{a}+(1-\zeta)\theta_{a}^2$ for $h_a=h_b$, where
$\theta_{a}$ is the frequency for haplotype $h_a$, and $\zeta$,
the fixation index quantifying the departure from HWE, is set to
0.05. We can see from the results listed in Table
\ref{Tab:violation result} that, with imputed genotype data, the
prospective-score test, like its exact-data counterpart, still
shows greater robustness in maintaining the type-I error rates
than the retrospective-score test. In particular, the
retrospective-score test, based on the recessive or dominant
model, may yield high type-I error rates under violation of HWE,
no matter whether exact or imputed genotype data are used. Thus,
an empirical-Bayes type shrinkage method that can adapt between
prospective and retrospective methods depending on bias-variance
trade-off could be useful for analysis of both typed and untyped
SNPs.

\begin{table}
\caption{Size (\%) of the prospective- and retrospective-score tests
(significance $\textit{level}=0.01$) based on the imputed and true (in
parenthesis) genotypes at the untyped causal SNP, using SNP data
generated according to Scenarios 1 (Table \protect\ref{Tab:scenario 1}) and
2 (Table \protect\ref{Tab:scenario 2}) and a fixation index of 0.5
(violating HWE). Results are based on 3000 simulated data sets}\label{Tab:violation result}
\begin{tabular*}{\columnwidth}{@{\extracolsep{\fill}}lcc@{}}
\hline
& \multicolumn{1}{c}{\textbf{Prospective score}} & \multicolumn
{1}{c@{}}{\textbf{Retrospective score}} \\
& \multicolumn{1}{c}{\textbf{imputed (true)}} &
\multicolumn{1}{c@{}}{\textbf{imputed (true)}} \\
\hline
\multicolumn{1}{@{}l}{Recessive model}&&\\
Scenario 1 & 0.8 (0.8) & 1.7 (1.7)\\
Scenario 2 & 1.2 (1.2) & 5.9 (7.7)\\[5pt]
\multicolumn{1}{@{}l}{Dominant model} &&\\
Scenario 1 & 0.9 (0.9) & 1.4 (1.4) \\
Scenario 2 &1.0 (0.8) & 3.2 (5.1)\\[5pt]
\multicolumn{1}{@{}l}{Additive model}&&\\
Scenario 1 & 1.0 (1.0) & 1.0 (1.0)\\
Scenario 2 & 0.7 (0.8) & 0.7 (0.8)\\
\hline
\end{tabular*}
\end{table}

We conclude this section with a discussion on the two types of
association analyses recently developed for untyped SNPs: the full
likelihood approach\break \citep{Lin2008AJHG} and the two-stage
approach
(\citeauthor{Nicolae2006GeneticEpi}, \citeyear{Nicolae2006GeneticEpi};
\citeauthor{Marchini2007NatureGenetics}, \citeyear{Marchini2007NatureGenetics}). The
full likelihood approach uses a retrospective likelihood for the
case-control sample and a likelihood for the external (such as
HapMap) data, by which the imputation and association analysis are
simultaneously performed in a one-stage manner. Conversely, the
two-stage approach performs the imputation and association
analysis separately: imputing missing genotypes in the first stage
and then performing association analysis in the second stage. In
the imputation stage of the two-stage approach, one can apply
existing powerful external imputation algorithms such as
\citeauthor{Nicolae2006GeneticEpi} (\citeyear{Nicolae2006GeneticEpi}) and
\citeauthor{Marchini2007NatureGenetics} (\citeyear{Marchini2007NatureGenetics}), and, hence, the two-stage
approach is convenient to implement. There has been some debate on
the efficiency difference between the two approaches
(\citeauthor{Marchini2008AJHG}, \citeyear{Marchini2008AJHG};
\citeauthor{Lin2008AJHGResponse}, \citeyear{Lin2008AJHGResponse}). Our simulation
results (Tables \ref{Tab:scenario 1 result} and \ref{Tab:scenario
2 result}) suggest that some of the efficiency difference between
the full likelihood and the two-stage approaches may be due to the
use of different likelihoods (prospective vs. retrospective) and
not so much due to the use of one-stage
vs. two-stage analysis. In this section we have
shown that one can still use a retrospective likelihood even in a
two-stage approach with powerful imputation performed at the first
stage.
\section{Haplotypes}\label{sec:sec4}

\subsection{Definitions, Background and Missing Data}
Although single-SNP association tests are often the primary
methods for genome-wide association scans, many secondary or
``downstream'' analyses are often useful for detailed
characterization of the risk of the disease associated with
specific genomic regions of interest. One popular technique is
\textit{haplotype-based association analysis}, which involves
studying the association of a disease with a genomic region in
terms of the underlying ``haplotypes,'' the combination of alleles
at multiple loci along individual homologous chromosomes.
Originally, haplotype-based association analysis was considered a
powerful technique for ``indirect'' association testing in
situations where a causal SNP may not have been genotyped, but the
haplotypes defined by multiple typed SNPs could serve as a good
``surrogate'' for the causal variant. With the advent of various
imputation methods, although haplotype analysis has become less
relevant for such indirect association testing, it remains a
useful tool for parsimonious characterization of disease risk
associated with multiple, possibly interacting, loci within a
given region. Moreover, it is conceivable that for some regions,
the haplotypes, and not the individual SNPs, are functional units
and, thus, for these regions stronger signals of associations could
be detected by performing haplotype-based regression analysis.

A technical problem for haplotype-based regression analysis is
that typically the haplotype information for the study subjects
is not directly observable. Instead, locus-specific genotype data
are observed, which contain information on the pair of alleles a
subject carries, but does not provide the ``phase information,''
that is, which combinations of alleles appear across multiple loci
along the individual chromosomes. In general, the genotype data
of a subject will be phase-ambiguous whenever the subject is
heterozygous at two or more loci. Statistically, the lack of phase
information can be viewed as a special missing data problem.

For example, suppose A/a and B/b denote the major/minor alleles in
two bi-allelic loci. A particular haplotype pair, called a
diplotype, is the pair of alleles that are inherited from one's
parents. One such haplotype pair would be $(AB)-(ab)$, and disease
risk can be associated with the number of copies of particular
haplotypes that one inherits. Unfortunately, the diplotypes are
not observable directly, but instead we can observe only the
unordered or combined genotypes, in this case $(Aa)$ at the first
locus and $(Bb)$ at the second locus, that is, $(AaBb)$. However,
when observing only the genotypes, the actual haplotype pair is
unknown, or ``phase ambiguous,'' because the haplotype pair
$(Ab)-(aB)$ has the same set of unordered genotypes. Confronted
with the unordered set of genotypes $(AaBb)$, we know that the
actual haplotype pair is either $(AB)-(ab)$ or $(Ab)-(aB)$, but we
must use probability models to take into account the phase
ambiguity when performing statistical inference.

In Section \ref{sec:sec2} we described ``model-free'' prospective
and ``model-based'' efficient retrospective methods for analyzing
SNP data, and we also described empirical-Bayes methods that
data-adaptively move between the two. Just as in SNP data, for
haplotype data there are also model-free and model-based methods,
and accompanying empirical-Bayes methods.

A variety of methods have been developed for\break haplotype-based
analysis of case-control data using the logistic regression model
(\citeauthor{Zhao2003AJHG}, \citeyear{Zhao2003AJHG};
\citeauthor{Lake2003HumanHeridity}, \citeyear{Lake2003HumanHeridity};
\citeauthor{Epstein2003AJHG}, \citeyear{Epstein2003AJHG};
\citeauthor{Satten2004GenetEpi}, \citeyear{Satten2004GenetEpi};
Spinka, Carroll and Chatterjee, \citeyear{Spinka2005GeneticEpi};
\citeauthor{Lin2006JASA}, \citeyear{Lin2006JASA};
\citeauthor{Chatterjee2006JASAcomment}, \citeyear{Chatterjee2006JASAcomment};
\citeauthor{Chen2009JASA}, \citeyear{Chen2009JASA}).
Consider a general risk model similar to (\ref{eq:logistic}) but
with the addition of environmental factors $(E)$ and written in
terms of the diplotypes, denoted as $H^{\mathrm{di}}$:
\begin{eqnarray}\label{eq:logistic2}
&&\operatorname{pr}(D=1|H^{\mathrm{di}},E)
\nonumber
\\[-8pt]
\\[-8pt]
\nonumber
&&\quad{}=\frac{\exp\{\alpha+m(H^{\mathrm{di}},E,\beta)\}}{1+\exp\{\alpha+m(H^{\mathrm{di}},E,\beta)\}},
\end{eqnarray}
where the function $m(\cdot)$ is chosen in a suitable way to reflect
an assumed mode of genetic effect. For example, suppose we are
interested in the particular haplotype $h_* = (ab)$. A model that
assumes an additive effect of this haplotype would have $m(H^{\mathrm{di}}=h^{\mathrm{di}},E)$ linear in the number of copies of the haplotype
$h_*$.

\subsection{Model-Based and Model-Free Methods}\label{subsec:3.2}

\def\di{^{\mathrm{di}}}

\subsubsection{Identifiability}
The data setup then is that we have observations on environmental
exposure $(E)$, genotypes $G$ and cases and controls $D$. What is
missing is the underlying diplotype $H^{\mathrm{di}}$. The
retrospective likelihood is still (\ref{eq:RL}), but the risk of
disease depends on the diplotype $H^{\mathrm{di}}$ and not otherwise on
the genotype.

While models such as (\ref{eq:logistic2}) seem straightforward
enough for random samples, in retrospective samples a problem
arises because of the phase ambiguity. In particular, all
components of $\beta$ may not be identifiable if the distribution
of $(H^{\mathrm{di}},E)$ is left completely unrestricted
(\citeauthor{Epstein2003AJHG}, \citeyear{Epstein2003AJHG}; \citeauthor{Lin2006JASA}, \citeyear{Lin2006JASA}). Thus, to make progress,
some type of distributional assumptions are needed. Here we will
distinguish between two approaches, both of them retrospective
in nature but with different distributional assumptions. The first
we call ``model-free'' in that very little is actually assumed
about the haplotype distribution. If haplotypes were observable,
this method reduces to ordinary prospective logistic regression,
while in the rare disease case with phase ambiguity, the method
reduces to that of \citeauthor{Zhao2003AJHG} (\citeyear{Zhao2003AJHG}). The second approach,
which we call ``model-based,'' makes much stronger assumptions
about the haplotype distribution, and reduces to the efficient
retrospective approach of \citeauthor{Chatterjee2005Biometrika} (\citeyear{Chatterjee2005Biometrika}) if
haplotypes were observable. The model-free method will thus be
more robust but less efficient than the model-based method.

\subsubsection{Model-based method}\label{subsec:4.2.2}

The model-based\break method \citep{Spinka2005GeneticEpi} has three aspects:
\begin{enumerate}[(A.1)]
\item[(A.1)] Haplotypes and the environment are
assumed independent in the population.
\item[(A.2)]
The diplotypes are assumed to be in HWE in the population, so that
\begin{eqnarray*}
&&\pr\bigl(H^{\mathrm{di}} = h\di=(h_a,h_b) \vert E\bigr)\\
&&\quad{}=q\{h\di=(h_a,h_b),\theta\}\\
&&\quad{}= \cases{\theta_a^2, & $\mbox{if }  h_a=
h_b$,\cr
2\theta_{a}\theta_{b}, & $\mbox{if }
h_a\neq h_b,$}
\end{eqnarray*}
where $\theta_s$ denotes the population frequency
for the haplotype $h_s$.
\item[(A.3)] The
distribution of the environmental variable $E$ is left completely
nonparametric.
\end{enumerate}

The methodology \citeauthor{Spinka2005GeneticEpi} (\citeyear{Spinka2005GeneticEpi}) used to construct their
profile likelihood was a nonparametric maximum likelihood estimator
over the unknown distribution of $E$. However, there is an
alternative derivation, one that is both more intuitive and much
easier to work out. Indeed, it is a not sufficiently well-known fact
that for most purposes a case-control study can be viewed as a
prospective study with missing data. Consider a sampling scenario
where each subject from the underlying population is selected into
the case-control study using a Bernoulli sampling scheme where the
selection probability for a subject given his/her disease status
$D=d$ is proportional to $N_d/\operatorname{pr}(D=d)$. Inference with the
actual case-control data can then be based on the pseudo-likelihood
derived for such an alternative sampling scenario. Let $\delta=1$
denote that a subject is selected in the case-control sample under
this Bernoulli sampling scheme and hence has been observed. Then in
this alternative sampling scheme, and with the assumptions stated
above,\break \citeauthor{Spinka2005GeneticEpi} (\citeyear{Spinka2005GeneticEpi}) compute\break $\pr(D=1, G=g \vert E,
\delta= 1)$. This calculation is simple and in the rare disease
case the resulting efficient model-based likelihood function reduces
to
\begin{eqnarray}\label{eq:rtext02}
 \hspace*{24pt}&&L_{\mathrm{model}}(D,G,E,\Omega)\nonumber\\
&&\quad= \sum_{h\di\in\H_G} q(h\di,\theta)
\exp[D\{\kappa+ m(h\di,E,\beta)\}]
\nonumber
\\[-8pt]
\\[-8pt]
\nonumber
&&\qquad{}\bigg/ \Biggl(\sum_{s=0}^1\sum_{h\di} q(h\di,
\theta) \\
&&\qquad{}\hspace*{45pt}\cdot\exp[s\{\kappa+ m(h\di,E,\beta)\}] \Biggr), \nonumber
\end{eqnarray}
where $p_d = \n_d/\n$, $\pi_d = \pr(D=d)$, $\kappa= \alpha+\break
\log(p_1/p_0) - \log(\pi_1/\pi_0)$, $\Omega=
(\beta,\theta,\kappa)$, and ${\H}_{G}$ is the set of diplotypes
consistent with the observed genotypes $G$.

\subsubsection{Model-free method}\label{subsec:4.2.3}

The two important model assumptions in the model-based estimator
are (A.1) and (A.2). Although because of identifiability some
model assumptions must be made, they can be weakened tremendously,
as follows (Chen, Chatterjee and Carroll, \citeyear{Chen2009JASA}):
\begin{enumerate}[(B.1)]
\item[(B.1)] The haplotype and the environment are
independent in the population given the genotype $G$.
\item[(B.2)] There population distribution
for the diplotypes given the genotype $G$, called $q_{\mathrm{free}}(h\di\vert G,\theta)$, can be derived assuming HWE.
\end{enumerate}
Following the same alternative sampling scheme as described in
Section \ref{subsec:4.2.2}, or by doing a nonparametric maximum
likelihood analysis, we can compute $\pr(D=1 \vert G, E,\delta=1)$
under assumptions (B.1), (B.2) and (A.3) to be
\begin{eqnarray} \label{eq:rtext02}
\qquad&&L_{\mathrm{free}}(D,G,E,\Omega)\nonumber\\
&&\quad{}=  \sum_{h\di\in\H_G} q_{\mathrm{free}}(h\di\vert G,\theta)\nonumber\\
&&\quad{}\hspace*{37pt}\cdot\exp[D\{\kappa+ m(h\di,E,\beta)\}]\\
&&\qquad{}\bigg/ \Biggl(\sum_{s=0}^1\sum_{h\di\in\H
_G} q_{\mathrm{free}}(h\di\vert G, \theta)\nonumber\\
&&\qquad{}\hspace*{58pt}\cdot \exp[s\{\kappa+ m(h\di,E,\beta)\}
] \Biggr).\nonumber
\end{eqnarray}
To see why the likelihood $L_{\mathrm{free}}$ requires far weaker
assumptions than $L_{\mathrm{model}}$, note that $L_{\mathrm{free}}$ requires
the haplotype--environment independence and HWE assumption only to
specify the conditional distribution $\operatorname{pr}(H^{\mathrm{di}}|G,X),$ while
$L_{\mathrm{model}}$ requires the same assumption to specify the entire
joint distribution $\operatorname{pr}
(H^{\mathrm{di}},X)$. As a result, $L_{\mathrm{free}}$ requires the haplotype--environment independence and HWE only
to resolve the phase ambiguous genotypes. The likelihood
contribution for the subjects with phase unambiguous genotypes, that is,
$G=H^{\mathrm{di}}$, is the same as that for the standard prospective
logistic regression. In contrast, $L_{\mathrm{model}}$ depends on the
assumptions (A.1) and (A.2) irrespective of whether a subject has a
missing phase
or not.

Note that $L_{\mathrm{free}}(D,G,E,\Omega)$ will contain little
information on $\theta$ since it conditions on $G$. Thus, when
implementing methods based on this likelihood,\break \citeauthor{Chen2009JASA} (\citeyear{Chen2009JASA})
proposed to replace the score function for~$\theta$ by the
estimating function for~$\theta$ based on the genotype data from
the controls and assuming that the haplotypes are in HWE in the
population.

\subsection{Empirical-Bayes}\label{subsec:4.3}

In Section \ref{subsec:4.2.2} we constructed a profile likelihood
under strong assumptions leading to an efficient method that will
not be robust to violations of the two major assumptions.
Conversely, in Section \ref{subsec:4.2.3} we computed a profile
likelihood leading to much more robust inference, but at the cost
of a steep loss of efficiency. Similarly to Section
\ref{subsec:2.3}, here we briefly review a fully sample size- and
data-adaptive empirical-Bayes method that\break \citeauthor{Chen2009JASA} (\citeyear{Chen2009JASA})
described for gaining efficiency when warranted but is still
robust.

Let $\wh{\beta}_{\mathrm{free}}$ and $\wh{\beta}_{\mathrm{model}}$ be the
model-free and model-based estimates, with $j\mathrm{th}$ components
$\wh{\beta}_{j,{\mathrm{free}}}$ and $\wh{\beta}_{j,{\mathrm{model}}}$. Let $V$
be the covariance matrix of $\wh{\psi} = \wh{\beta}_{\mathrm{free}} -
\wh{\beta}_{\mathrm{model}}$, with the $j\mathrm{th}$ diagonal element of $V$
being $v_j$: a sandwich estimator $v_j$ can be computed, although
a nonparametric bootstrap can also be used. Then one can define
the empirical-Bayes estimator
\begin{eqnarray}
\wh{\beta}_{j,{\mathrm{\mathit{EB}}}} &=&
\wh{\beta}_{j,{\mathrm{free}}} + W_j
(\wh{\beta}_{j,{\mathrm{model}}}-\wh{\beta}_{j,{\mathrm{free}}}); \label{eq:qrjc05}
\nonumber
\\[-8pt]
\\[-8pt]
\nonumber
W &=& \frac{v_j}{v_j + (\wh{\beta}_{j,{\mathrm{free}}}-\wh{\beta}_{j,{\mathrm{model}}})^2}.
\end{eqnarray}
The intuition behind (\ref{eq:qrjc05})
is that if the model fails, $(\wh{\beta}_{j,{\mathrm{model}}}-\wh{\beta}_{j,{\mathrm{free}}})$ will be large relative to $v_j$,
which as a variance is proportional to $N^{-1}$, hence, $W_j
\approx0$, and, hence, the empirical-Bayes method will effectively
become the model-free estimator. If, however, the model assumption
holds, then $v_j$ and $(\wh{\beta}_{j,{\mathrm{free}}}-\wh{\beta}_{j,{\mathrm{model}}})^2$ are proportional to one another, so that $W_j > 0$ and
the empirical-Bayes estimate goes part way toward the model-based
estimator, and hence gains efficiency over the model-free
estimate.\break \citeauthor{Chen2009JASA} (\citeyear{Chen2009JASA}) describe the\break limiting distribution
of (\ref{eq:qrjc05}) and how to compute an estimate of its
variance.

\citeauthor{Chen2009JASA} (\citeyear{Chen2009JASA}) illustrate application of the different methods
in two case-control data examples. The examples were chosen in such a
way that from a priori biologic grounds one would expect the
gene--environment independence assumption to hold in one case, but not
in the other. The two examples together illustrate how the
different shrinkage estimators adapt to alternative scenarios of
gene--environment distribution.

\section{Discussion}
Researchers now increasingly use the Cochran--Armitage trend test
as the primary method for single-SNP association testing in the
GWAS. The test is known to have robust power for the detection of
effect of susceptibility SNPs under a range of realistic modes of
inheritance that give rise to some sort of monotone relationship
between disease risk and allele count. As noted in Section \ref{sec:sec2}, the
retrospective and prospective methods have very similar, if not
identical, power under the trend model and thus either could be
used as the primary method for analysis of GWAS data. The trend
test, however, can perform very poorly for the detection of SNPs
for which the minor allele has a recessive effect. Thus, it is
often recommended that a test under the recessive mode of
inheritance be conducted as a secondary step to detect SNPs with
recessive effects that may be missed by the primary trend test of
association. The use of the retrospective method can be
potentially beneficial at this stage. One, however, has to be
cautious about creation of false positive results due to the
violation of the HWE assumption. We recommend that if a
retrospective method is to be used for potential power gain, then
it should be used in conjunction with the empirical-Bayes type
shrinkage estimation. Our numerical investigations suggest that
such a method can indeed be more powerful than the conventional
``prospective'' methods without creating excess false positives;
see Tables \ref{Tab:observedPval} and \ref{Tab:rank_permute}.

In this article, although we focus on association tests involving
bi-allelic SNPs, the same issues are relevant for genetic
association tests involving loci with more than two alleles. In
particular, one can gain efficiency for analysis of case-control
data by assuming HWE or other natural population-genetic models
(\citeauthor{Satten2004GenetEpi}, \citeyear{Satten2004GenetEpi};
\citeauthor{Lin2006JASA},\break \citeyear{Lin2006JASA}) to specify multi-allelic
genotype frequency for the underlying population. The sensitivity of
the methods to underlying model assumption can be reduced by
appropriate shrinkage estimation techniques.

The impact of population stratification (PS) can be very different for
prospective and retrospective methods. As it is well known, the
presence of\break \mbox{population} stratification, that is, the existence of hidden
ethnic sub-structures in
the population, can create confounding bias in all of the methods when
both gene-frequency and disease risks vary across the \mbox{underlying}
strata. The presence of PS can also cause large scale violation of the
HWE assumption, thus making the retrospective method more susceptible
to bias than its prospective counterpart. Our application of different
methods to the CGEMS\break genome-wide association study data illustrated
that the empirical-Bayes type procedure can correct for inflated type-I
error that may exist for retrospective methods due to large scale
violation of the underlying HWE assumption.

The difference between prospective and retrospective methods
becomes more relevant for studies of gene--gene and
gene--environment interactions, a topic that we have not directly
addressed in this article. In particular, retrospective methods,
such as the case-only analysis \citep{Piegorsch1994StatMed},
which assumes gene--gene or/and gene--environment independence for
the underlying population, can gain dramatic power for testing and
estimation of odds ratio interaction parameters in the logistic
regression model. Given that standard case-control analyses often
have poor power for detection of multiplicative interactions due
to small numbers of cases or controls in cells of crossing
exposures, practitioners often find it is tempting to use the more
powerful retrospective methods. The assumption of gene--environment
independence, however, can be violated, either due to direct
casual association between gene and environment or indirect
association due to effects of family history and hidden population
stratification. The assumption of gene--gene independence between
physically distant genes can also be violated due to population
stratification. Thus, we believe the development of shrinkage
(Mukherjee and Chatterjee, \citeyear{Mukherjee2008Biometrics}; Chen, Chatterjee and Carroll,
\citeyear{Chen2009JASA}) and other types of data-adaptive
techniques\break
(\citeauthor{Li2009AJE}, \citeyear{Li2009AJE}) has been valuable for robust inference in
case-control studies of genetic epidemiology.

\section*{Acknowledgments}
Chatterjee's research was supported by a gene--environment
initiative grant from the National Heart Lung and Blood Institute
(RO1HL091172-01) and by the Intramural Research Program of the
National Cancer Institute. Chen's research was supported by the
National Science Council of ROC (NSC 95-2118-M-001-022-MY3).
Carroll's research was supported by a grant from the National
Cancer Institute\break (CA57030) and by Award Number KUS-CI-016-04, made
by King Abdullah University of Science and Technology (KAUST).


\begin{thebibliography}{}

\bibitem[\protect\citeauthoryear{Andersen}{1970}]{Andersen1970JRSSB}
\textsc{Andersen, E. B.} (1970).
 Asymptotic properties of conditional maximum-likelihood estimators.
\textit{J. Roy. Statist. Soc. Ser. B} \textbf{32} 283--301.
\MR{0273723}

\bibitem[\protect\citeauthoryear{Chapman et al.}{2003}]{Chapman2003HumanHeredity}
\textsc{Chapman, J. M., Cooper, J. D., Todd, J. A.} and \textsc{Clayton,~D. G.} (2003).
 Detecting disease associations due to linkage disequilibrium using
haplotype tags: A class of tests and the determinants of statistical power.
\textit{Human Heredity} \textbf{56} 18--31.

\bibitem[\protect\citeauthoryear{Chatterjee and Carroll}{2005}]{Chatterjee2005Biometrika}
\textsc{Chatterjee, N.} and \textsc{Carroll, R. J.} (2005).
 Semiparametric maximum likelihood estimation exploiting
gene--environment independence in case-control studies.
\textit{Biometrika} \textbf{92} 399--418.
\MR{2201367}

\bibitem[\protect\citeauthoryear{Chatterjee et al.}{2006}]{Chatterjee2006JASAcomment}
\textsc{Chatterjee, N., Spinka, C., Chen, J.} and \textsc{Carroll, R. J.} (2006).
 Likelihood based inference on haplotype effects in genetic
association studies-{C}omment.
\textit{J. Amer. Statist. Assoc.} \textbf{101} 108--111.

\bibitem[\protect\citeauthoryear{Chen and Chatterjee}{2007}]{Chen2007HumHered}
\textsc{Chen, J.} and \textsc{Chatterjee, N.} (2007).
 Exploiting {H}ardy--{W}einberg equilibrium for efficient
screening of
single {SNP} associations from case-control studies.
\textit{Human Heredity} \textbf{63} 196--204.

\bibitem[\protect\citeauthoryear{Chen, Chatterjee and Carroll}{2009}]{Chen2009JASA}
\textsc{Chen, Y. H., Chatterjee, N.} and \textsc{Carroll, R. J.} (2009).
 Shrinkage estimators for robust and efficient inference in
haplotype-based case-control studies.
\textit{J. Amer. Statist. Assoc.} \textbf{104} 220--233.

\bibitem[\protect\citeauthoryear{Cornfield}{1956}]{Cornfield1956}
\textsc{Cornfield, J.} (1956).
 A statistical problem arising from retrospective studies.
In \textit{Proceedings of the Third Berkeley Sympos. Math.
Statist. Probab.} 135--148. Univ. California Press, Berkeley.
\MR{0084935}

\bibitem[\protect\citeauthoryear{Epstein and Satten}{2003}]{Epstein2003AJHG}
\textsc{Epstein, M. P.} and \textsc{Satten, G. A.} (2003).
 Inference on haplotype effects in case-control studies using unphased
genotype data.
\textit{American Journal of Human Genetics} \textbf{73} 1316--1329.

\bibitem[\protect\citeauthoryear{Hartl and Clark}{2007}]{Hartl2007}
\textsc{Hartl, D. L.} and \textsc{Clark, A. G.} (2007).
\textit{Principles of Population Genetics}, 4th ed.
 Sinauer Associates, Sunderland, MA.

\bibitem[\protect\citeauthoryear{Hunter et al.}{2007}]{Hunter2007NatGenet}
\textsc{Hunter, D. J., Kraft, P., Jacobs, K. B., Cox, D. G., Yeager, M., Hankinson,
S. E., Wacholder, S., Wang, Z., Welch, R., Hutchinson, A., Wang, J.,
Yu, K.,
Chatterjee, N.} et al. (2007).
 A genome-wide association study identifies alleles in {FGFR}2
associated with risk of sporadic postmenopausal breast cancer.
\textit{Nature Genetics} \textbf{39} 870--874.

\bibitem[\protect\citeauthoryear{Lake et al.}{2003}]{Lake2003HumanHeridity}
\textsc{Lake, S. L., Lyon, H., Tantisira, K., Silverman, E. K., Weiss, S. T., Laird,
N. M.} and \textsc{Schaid, D. J.} (2003).
 Estimation and tests of haplotype--environment interaction when
linkage phase is ambiguous.
\textit{Human Heredity} \textbf{55} 56--65.

\bibitem[\protect\citeauthoryear{Li and Conti}{2009}]{Li2009AJE}
\textsc{Li, D.} and \textsc{Conti, D. V.} (2009).
 Detecting gene--environment interactions using a combined case-only
and case-control approach.
\textit{American Journal of Epidemiology} \textbf{169} 497--504.

\bibitem[\protect\citeauthoryear{Lin and Hu}{2008}]{Lin2008AJHGResponse}
\textsc{Lin, D. Y.} and \textsc{Hu, Y.} (2008).
 Reply to {M}archini and {H}owie.
\textit{American Journal of Human Genetics} \textbf{83} 539--540.

\bibitem[\protect\citeauthoryear{Lin, Hu and Huang}{2008}]{Lin2008AJHG}
\textsc{Lin, D. Y., Hu, Y.} and \textsc{Huang, B. E.} (2008).
 Simple and efficient analysis of disease association with missing
genotype data.
\textit{American Journal of Human Genetics} \textbf{82} 444--445.

\bibitem[\protect\citeauthoryear{Lin and Zeng}{2006}]{Lin2006JASA}
\textsc{Lin, D. Y.} and \textsc{Zeng, D.} (2006).
 Likelihood-based inference on haplotype effects in genetic
association studies.
\textit{J. Amer. Statist. Assoc.}
\textbf{101} 89--104.
\MR{2268031}

\bibitem[\protect\citeauthoryear{Luo et al.}{2009}]{Luo2009GeneticEpi}
\textsc{Luo, S., Mukherjee, B., Chen, J.} and \textsc{Chatterjee, N.} (2009).
 Shrinkage estimation for robust and efficient screening of
single-{SNP} sssociation from case-control genome-wide association studies.
\textit{Genetic Epidemiology} Online.

\bibitem[\protect\citeauthoryear{Marchini and Howie}{2008}]{Marchini2008AJHG}
\textsc{Marchini, J.} and \textsc{Howie, B.} (2008).
 Comparing algorithms for genotype imputation.
\textit{American Journal of Human Genetics} \textbf{83} 535--539.

\bibitem[\protect\citeauthoryear{Marchini et al.}{2007}]{Marchini2007NatureGenetics}
\textsc{Marchini, J., Howie, B., Myers, S., McVean, G.} and \textsc{Donnelly, P.} (2007).
 A new multipoint method for genome-wide association studies by
imputation of genotypes.
\textit{Nature Genetics} \textbf{39} 906--913.\

\bibitem[\protect\citeauthoryear{Mukherjee and Chatterjee}{2008}]{Mukherjee2008Biometrics}
\textsc{Mukherjee, B.} and \textsc{Chatterjee, N.} (2008).
 Exploiting gene--environment independence for analysis of case-control
studies: An empirical {B}ayes approach to trade off between bias and
efficiency.
\textit{Biometrics} \textbf{64} 685--694.

\bibitem[\protect\citeauthoryear{Nicolae}{2006}]{Nicolae2006GeneticEpi}
\textsc{Nicolae, D. L.} (2006).
 Testing untyped alleles ({TUNA})-applications to genome-wide
association studies.
\textit{Genetic Epidemiology} \textbf{30} 718--727.

\bibitem[\protect\citeauthoryear{Piegorsch, Weinberg and Taylor}{1994}]{Piegorsch1994StatMed}
\textsc{Piegorsch, W. W., Weinberg, C. R.} and \textsc{Taylor, J. A.} (1994).
 Non-hierarchical logistic models and case-only designs for assessing
susceptibility in population-based case-control studies.
\textit{Statist. Med.} \textbf{13} 153--162.

\bibitem[\protect\citeauthoryear{Prentice and Pyke}{1979}]{Prentice1979Biometrika}
\textsc{Prentice, R. L.} and \textsc{Pyke, R.} (1979).
 Logistic disease incidence models and case-control studies.
\textit{Biometrika} \textbf{66} 403--412.
\MR{0556730}

\bibitem[\protect\citeauthoryear{Roeder, Carroll and Lindsay}{1996}]{Roeder1996JASA}
\textsc{Roeder, K., Carroll, R. J.} and \textsc{Lindsay, B. G.} (1996).
 A~semiparametric mixture approach to case-control studies
with errors
in covariables.
\textit{J. Amer. Statist. Assoc.} \textbf{91} 722--732.
\MR{1395739}

\bibitem[\protect\citeauthoryear{Satten and Epstein}{2004}]{Satten2004GenetEpi}
\textsc{Satten, G. A.} and \textsc{Epstein, M. P.} (2004).
 Comparison of prospective and retrospective methods for haplotype
inference in case-control studies.
\textit{Genetic Epidemiology} \textbf{27} 192--201.

\bibitem[\protect\citeauthoryear{Satten and Kupper}{1993}]{Satten1993JASA}
\textsc{Satten, G. A.} and \textsc{Kupper, L. L.} (1993).
 Conditional regression analysis of the exposure-disease odds ratio
using known probability-of-exposure values.
\textit{Biometrics} \textbf{49} 429--440.

\bibitem[\protect\citeauthoryear{Spinka, Carroll and Chatterjee}{2005}]{Spinka2005GeneticEpi}
\textsc{Spinka, C., Carroll, R. J.} and \textsc{Chatterjee, N.} (2005).
 Analysis of case-control studies of genetic and environmental factors
with missing genetic information and haplotype-phase ambiguity.
\textit{Genetic Epidemiology} \textbf{29} 108--127.

\bibitem[\protect\citeauthoryear{Thomas et al.}{2008}]{Thomas2008NatGenet}
\textsc{Thomas, G., Jacobs, K. B., Yeager, M., Kraft, P., Wacholder, S., Orr,
N., Yu,
K., Chatterjee, N., Welch,~R., Hutchinson, A.} et al. (2008).
 Multiple novel loci identified in a genome-wide association
study of
prostate cancer.
\textit{Nature Genetics} \textbf{40} 310--315.

\bibitem[\protect\citeauthoryear{van Belle et al.}{2004}]{VanBelle2004BiostatisticsBook}
\textsc{van Belle, G., Heagerty, P. J., Fisher, L. D.} and \textsc{Lumley,~T. S.} (2004).
\textit{Biostatistics: A Methodology for the Health Sciences}.
 Wiley, Hoboken, NJ.

\bibitem[\protect\citeauthoryear{Xiong, Zhao and Berwinkle}{2002}]{Xiong2002AJHG}
\textsc{Xiong, M., Zhao, J.} and \textsc{Berwinkle, E.} (2002).
 Generalized {T}2 test for genome association studies.
\textit{American Journal of Human Genetics} \textbf{70} 1257--1268.

\bibitem[\protect\citeauthoryear{Yeager et al.}{2007}]{Yeager2007NatGenet}
\textsc{Yeager, M., Orr, N., Hayes, R. B., Jacobs, K. B., Kraft, P., Wacholder, S.,
Minichiello, M. J., Fearnhead, P., Yu, K., Chatterjee, N.} et al.
(2007).
 Genome-wide association study of prostate cancer identifies a second
risk locus at 8q24.
\textit{Nature Genetics} \textbf{39} 645--649.

\bibitem[\protect\citeauthoryear{Yu et al.}{2009}]{Yu2009GeneticEpi}
\textsc{Yu, K., Li, Q., Bergen, A. W., Pfeiffer, R., Rosenberg, P., Caporaso, N., Kraft,
P.} and \textsc{Chatterjee, N.} (2009).
 Pathway analysis by adaptive combination of {P}-values.
\textit{Genetic Epidemiology} \textbf{33} 700--709.

\bibitem[\protect\citeauthoryear{Zhao, Li and Khalid}{2003}]{Zhao2003AJHG}
\textsc{Zhao, L. P., Li, S. S.} and \textsc{Khalid, N. A.} (2003).
 Method for the assessment of disease associations with
single-nucleotide polymorphism haplotypes and environmental variables in
case-control studies. \textit{American Journal of Human Genetics} \textbf{72} 1231--1250.

\end{thebibliography}
\end{document}